\documentclass[preprint]{aastex}
\usepackage{longtable,lscape}
\usepackage{natbib}
\usepackage[dvips]{color}
\usepackage{multirow}
\newcommand{\HII}{H\,{\sc ii}}

\newcommand{\LZ}{{\it L-Z}}
\newcommand{\OIII}{[O\,{\sc iii}]\,$\lambda$}
\newcommand{\OII}{[O\,{\sc ii}]\,$\lambda$}
\begin{document}
\shorttitle{ {\it L-Z} Relations for BCDs in the optical and NIR}
\shortauthors{Zhao, Gao \& Gu}
\title{Luminosity-Metallicity Relations for Blue Compact Dwarf Galaxies in the Optical and Near-Infrared}
\author{Yinghe Zhao and Yu Gao}
\affil{Purple Mountain Observatory, Chinese Academy of Sciences (CAS), Nanjing 210008, China}
\email{yhzhao, yugao@pmo.ac.cn}
\and
\author{Qiusheng Gu}
\affil{Department of Astronomy, Nanjing University, Nanjing 210093, China}
\email{qsgu@nju.edu.cn}

\date{Received:~ Accepted:~}
\begin{abstract}
In this paper, we present systematic studies on the  {\it B}-, {\it R}- and $K_s$-band luminosity-metallicity (\LZ) relations for a set of metal poor, blue compact dwarf galaxies. Metallicity is derived by using both the empirical N2 and the direct $T_e$ methods. Our work reconciles contradictory results obtained by different authors and shows that the \LZ\ relationship does also hold for blue compact dwarf galaxies. The empirical N2-based slope of the \LZ\ relation, for each photometric band, is consistent with the $T_e$-based one. We confirm that the slope of the \LZ\ relation is shallower in the near-infrared than that in the optical. Our investigations on the correlations between the {\it $L_B$-Z} relation residuals and different galactic parameters show that the star formation activities could be a cause of the large scatter in the optical \LZ\ relationships, whereas the internal-absorption might be another possible contributing factor.
\end{abstract}

\section{Introduction}
Since the seminal work of Lequeux et al. (1979), the relationship between luminosity ({\it L}) and metallicity ({\it Z}; hereafter \LZ\ relation), two crucial parameters for understanding the behavior of galaxies, has been extensively studied (e.g. Skillman et al. 1989; Tremonti et al. 2004). Metallicity reflects the gas reprocessed by stars and any exchange of gas between environments and galaxies, while stellar mass (i.e. luminosity for constant mass-to-light ratio) reflects the amount of gas locked up into stars. Therefore, the relationship between luminosity and metallicity can provide an important constraint on the models of galaxy formation and evolution that attempt to account for the chemical evolution of the system (see e.g., Prantzos \& Boissier 2000; Boissier et al. 2003; Qian \& Wasserburg 2004). Additionally, the \LZ\ relation could be also used to search for the most metal-poor galaxies and to derive, at least to a first approximation, the distance of a galaxy from its metallicity and vice versa. 

A well-defined correlation between the blue luminosity ($L_B$) and the metallicity for dwarf irregulars (dIs), spirals and ellipticals has been established by various authors (e.g., Garnett \& Shields 1987; Skillman et al. 1989; Zaritsky et al. 1994; Melbourne \& Salzer 2002; Lee et al. 2003; Lamareille et al. 2004; Tremonti et al. 2004; Salzer et al. 2005; van Zee \& Haynes 2006), over $\sim$10 mag in luminosity and 2 dex in metallicity. However, some studies do not support these results for all types of galaxies. Specially, several contradictory results regarding the existence of an \LZ\ relationship of dwarf galaxies have been presented over years. Hidalgo-G\'amez \& Olofsson (1998) presented a weaker \LZ\ relationship than previously thought, between $M_B$ and O/H for dIs. Hunter \& Hoffman (1999; hereafter HH99) found that the relationship between $M_B$ and O/H for Im, Sm and blue compact dwarf galaxies (BCDs) has a very large scatter. Another question is whether the \LZ\ relationship for dwarf galaxies (if it is true) exists in a similar manner as for massive galaxies. Using a sample of more than 500 star-forming galaxies, Melbourne \& Salzer (2002) found that the \LZ\ relation for giant galaxies has less scatter and is steeper than that for dwarf galaxies (also see Salzer et al. 2005). Using $\sim$1000 individual spectra of \HII\ regions in 54 late-type galaxies, however, Pilyugin et al. (2004) found that the slope of the \LZ\ relationship for spirals ($-18<M_B<-22$ mag) is slightly shallower than the one for dIs ($-12<M_B<-18$ mag), which is consistent with the result for blue compact galaxies (BCGs; Shi et al. 2005). 

BCDs are a group of extragalactic objects that are spectroscopically characterized by a faint, blue optical continuum accompanied, in most cases, by intense emission lines. BCDs are small galaxies with low metallicities ($1/50<Z<1/3\ Z_\odot$; HH99) and have dramatically different properties compared to normal dwarf galaxies (Zwicky 1966; Gil de Paz et al. 2003). These galaxies have bluer colors than ordinary dIs and their surface brightness is much higher. Accumulated observational evidence over the recent years provided more details on the unique properties of these galaxies (for a review see Kunth \& \"Ostlin 2000). Having a relatively low metallicity, BCDs are at an early epoch of their evolution, making them similar to samples of the distant, more massive protogalaxies, thus allowing us to study the star formation and chemical enrichment in an environment likely to be similar to that in the early universe.

Although it is thought the \LZ\ relationship should exist in almost all types of galaxies, it is not clear yet whether this relationship holds for BCDs alone (HH99, Fig. 7; Hopkins et al. 2002, Fig. 7). Vaduvescu et al. (2007) show that there exists a good correlation between the metallicity and the $K_s$ luminosity for BCDs. However, results from these authors may suffer large uncertainties from the limited size of their BCD sample (13, 22 and 14 galaxies in HH99, Hopkins et al. and Vaduvescu et al., respectively),  and/or the selection criteria for BCDs, as well as the inconsistence in the determination of metallicity for different objects. Therefore, further works are needed to disentangle these two apparently contradictory results.

In this paper, we present our systemic studies on the \LZ\ relationships for BCDs from the optical to the near-infrared (NIR), i.e. at the {\it B}-, {\it R}- and {\it $K_s$}-band. Our BCD sample is derived from the data in Gil de Paz et al. (2003), and it contains $\sim70$ galaxies (40 of which have NIR photometry) which is about 6 times larger than the one used in previous studies of dwarf galaxies. The NIR data are useful to study the \LZ\ relation because they are less affected by the ongoing star formation activities, and they also suffer significantly less from dust extinctions (both external and internal). Moreover, the $K_s$ luminosity is a more reliable measurement of the stellar mass in these galaxies (Gil de Paz 2000; P\'erez-Gonz\'alez et al. 2003). Hence, the study of $L_{K_s}$-$Z$ relationship allows us to tie the observed variations in metal abundance to stellar mass, one of the most fundamental physical parameters of galaxies. However, there are only few works (Salzer et al. 2005; Mendes de Oliveira et al. 2006; Vaduvescu et al. 2007) studying the \LZ\ relation in the NIR due to the lack of photometric data. 

The paper is organized as following: Section 2 describes the sample of BCDs and data reductions. In \S 3 we first derive the oxygen abundances for a subsample of BCD galaxies which have SDSS spectroscopical observations, and then we evaluate the {\it B-, R-} and {\it $K_s$}-band \LZ\ diagrams. We compare our results with previous work, discuss the slopes of our \LZ\ relations and explore the possible origins of the scatter in the \LZ\ relation in \S 4. In the last section, a brief summary is present. 

\section{Sample and Data Reduction}
There have been a number of criteria for selecting BCDs. These criteria are commonly based on the galaxy's morphological properties and its luminosity (Zwicky \& Zwicky 1971; Thuan \& Martin 1981), although some definitions are based on their spectroscopic properties (Gallego et al. 1996). BCDs classified by these different criteria may be confused with each other (Gil de Paz et al. 2003). However, for our purpose, i.e. to determine the \LZ\ relationship, a clear and homogeneous sample of blue compact dwarf galaxies is necessary.
 
Using a unified concept of BCD galaxy, Gil de Paz et al. (2003) compiled a BCD sample (including 104 members) from several exploratory studies. The galaxies in their sample were selected by putting forward a new set of quantitative classification criteria. This new set of criteria is a combination of the galaxy's color, morphology and luminosity. Briefly, a BCD galaxy has to fulfill the following three criteria (referring to {\it blue, compact} and {\it dwarf}, respectively): (1) $\mu_{B,\rm{peak}}-\mu_{R,\rm{peak}} < 1$, where $\mu_{B,\rm{peak}}$ and $\mu_{R,\rm{peak}}$ are the peak surface brightness of {\it B}- and {\it R}-band, respectively, (2) $\mu_{B,\rm{peak}} < 22$ mag arcsec$^{-2}$, and (3) the absolute magnitude $M_K > -21$ mag [see Gil de Paz et al. (2003) for details]. Such a large and universal sample allows us to measure an unbiased \LZ\ relation for BCDs. 

In this work, we select BCD galaxies from the Gil de Paz's sample, which is the largest well-defined sample of BCDs. Since not all galaxies have their metallicities measured, a galaxy is selected if (1) we can get its oxygen abundance from literature, or (2) its [N\,{\sc ii}]\,$\lambda6584$/H$_\alpha$ ratio was given in the original sample, or (3) we can obtain a spectrum from Sloan Digital Sky Survey (SDSS; Gunn et al. 1998; Blanton et al. 2003 ) Data Release Six (DR6).

To derive the oxygen abundances for galaxies which have SDSS spectra, we need to measure the fluxes of emission-lines accurately. Thus we have to correct both the Galactic reddening and the underlying stellar absorption. The detailed method for the latter process can be found in Zhao et al. (2009; in preparation). We only give a brief description here. First, the observed spectra were dereddened for Galactic extinction, using the extinction coefficients from Schlegel et al. (1998) and the empirical extinction law from Cardelli et al. (1989). Then we model the stellar contribution through the stellar population synthesis code, STARLIGHT {\scriptsize version 2.0} (Cid Fernandes et al. 2004). After subtracting the best-fit model spectrum from the observed one, we obtain the ``pure" emission-line spectrum, from which we could measure accurate fluxes for all emission lines with the {\it onedspec.splot} task in IRAF \footnote{IRAF is distributed by the National Optical Astronomical Observatory, which is operated by the Association of Universities for Research in Astronomy (AURA), Inc., under a cooperative agreement with the National Science Foundation.}. The line flux errors are typically less than 5\%. 

Finally, the observed line intensities were corrected for the intrinsic extinctions from the equation
\begin{equation}
\frac{I(\lambda)}{I({\rm{H}}\beta)}=\frac{F(\lambda)}{F({\rm{H}}\beta)} 10^{c({\rm{H}}\beta) f(\lambda)}
\end{equation}
where $I(\lambda)$ is the intrinsic line flux, $F(\lambda)$ is the observed line flux, $c({\rm{H}}\beta)$ is the logarithmic reddening parameter and $f(\lambda)$ is the redding function (Cardelli et al. 1989). For the intrinsic hydrogen line intensity ratios, we used the theoretical ratios from Brocklehurst (1971) at electron temperature and number density estimated from the observed \OIII4959,5007/$\lambda4363$ ratio and [S\,{\sc ii}]\,$\lambda6717$/$\lambda6731$ ratio, respectively (see \S 3.1 for detail).

The photometric data for the optical bands are adopted from Gil de Paz et al. (2003). The NIR data used here are mainly taken from the Two Micron All Sky Survey (2MASS) extended source catalog (Jarrett et al. 2000). For several objects the $K_s$ magnitudes are adopted from Noeske et al. (2003; 2005) and Vaduvescu et al. (2007). Because of the relatively shallower observations of 2MASS, we prefer using the data from the later two sources to using the data from 2MASS when overlapping sources are found.

\section{The \LZ\ Relationships}
\subsection{The Metallicity}
Generally, the O/H ratio is used to study the \LZ\ relationship since it is easily attained in the optical part of the spectrum. The preferred method for determining the oxygen abundance in galaxies is through electron temperature-sensitive lines (the so-called $T_e$ method), such as the \OIII4363 line (Aller 1984). The main problem is that normally the \OIII4363 line is very weak and only appears in very high excitation spectra. When the \OIII4363 line is absent, some other empirical methods, such as the $R_{23}$ method (Pagel et al. 1979), the $P$ method (Pilyugin 2001) and the N2 method (Denicol\'o et al. 2002) have been employed frequently. These methods do not need any direct measurement of the electron temperature. 

\subsubsection{$T_e$ Method}
We used the formula given by Izotov et al. (2006) to determine the abundances for galaxies which have good signal-to-noise ratios for the \OIII4363 line. Since the wavelength coverage of the SDSS spectra is 3800-9300 \AA, the \OII3727 lines are not available for the low redshift ($<0.02$) BCDs used here. Therefore, we used their equation (4), which makes use of the \OII7320, 7331 lines instead of the \OII3727 line, to calculate the (O$^+$/H). The formulae is given the form of
\begin{equation}
12 + {\rm{log(O^+/H^+}})={\rm{log}}\frac{\lambda7320+\lambda7331}{{\rm{H}}\beta}+6.901+\frac{2.487}{t}-0.483 {\rm{log}} t-0.013 t+{\rm{log}} (1-3.48x)
\end{equation}
where $t=10^{-4} T_e$[O\,{\sc iii}], and $x=10^{-4}n_e t^{-0.5}$. To calculate the (O$^{2+}$/H), we used the equation (5) in Izotov et al. (2006), as following
\begin{equation}
12 + {\rm{log(O^{2+}/H^+}})={\rm{log}}\frac{\lambda4959+\lambda5007}{{\rm{H}}\beta}+6.200+\frac{1.251}{t}-0.55 {\rm{log}} t-0.014 t
\end{equation}
The electron temperature ($T_e$) and number density ($n_e$) were derived with the IRAF task {\it nebular.zones} (De Robertis et al. 1987; Shaw \& Dufour 1995) using the \OIII4363/$(\lambda4959+\lambda5007)$ ratio and the [S\,{\sc ii}]\,$\lambda6717$/$\lambda6731$ ratio, respectively. This task is based on a five-level statistical equilibrium model and makes use of the latest collision strengths and radiative transition probabilities. For those we can not obtain the electron densities from the [S\,{\sc ii}] line ratio, we assumed a density of 100 cm$^{-3}$. This assumption almost does not affect our oxygen abundance determination, since the effect of temperature is much larger than that of electron density. For 12 out of 24 BCDs which have SDSS spectroscopical observations, we've measured the oxygen abundance using the $T_e$ method, which is presented in Table 1.

To study the \LZ\ relation, we also compiled oxygen abundance determined by the $T_e$ method from the literature, as shown in Column (6) of Table 2. When more than one data point for a galaxy can be found, we use the error-weighted average value. For galaxies we can not obtain errors of abundance from the original references, an equal weighted mean value is adopted. The final error is the bigger one of the two values: half of the difference of the range in values and the propagated error.

 \subsubsection{N2 Method}
 
The N2[$\equiv$\,log\,([N\,{\sc ii}]\,$\lambda$6584/H$_\alpha$)] empirical method is proposed by Denicol\'o et al. (2002), following the earlier work by Storchi-Bergmann et al. (1994) and Raimann et al. (2000). They used a representative sample of H\,{\sc ii} galaxies which have accurate oxygen abundances, plus photoionization models covering a wide range of abundances ($7.2<{\rm{log(O/H)}}<9.1$), to calibrate the N2 estimator. As shown in Denicol\'o et al. (2002), the oxygen abundance and the N2 calibrator are well correlated (the correlation coefficient is 0.85), and the linear relation holds for the whole metallicity range, from the most metal-poor to the most metal-rich galaxies in their sample. Least-squares fits to the data simultaneously minimizing the errors in both axes, give
\begin{equation}
12+\rm{log(O/H)}=9.12+0.73\times\rm{N}2.
\end{equation} 

The oxygen abundance determined based on this method has a precision of $\sim 0.2$ dex. The N2 parameter has clear observational advantages for ranking metallicities in star-forming galaxies. Besides it being monotonic with log(O/H), it is also independent on reddening correction and flux calibration. What's more, for the BCDs having SDSS spectroscopic observations, the \OII3727 line, which is critical both for the $R_{23}$ method and the $P$ method, is not available in the spectra. In addition to the above reasons, we have another reason for using N2 method other than the $P$ or the $R_{23}$ method to derive the oxygen abundances for the BCD sample used in the current work. The [N\,{\sc ii}]/H$_\alpha$ ratio of about 47\% galaxies were given by Gil de Paz et al. (2003), thus this method allows us to collect more data points than the other methods. The oxygen abundances determined based on the N2 method are listed in Table 1 (for the SDSS sample) and Column (7) of Table 2 (for the entire sample). 

\subsection{The \LZ\ Relations in the {\it B-} and {\it R}-band}
In Table 2 we list the {\it B} and {\it R} magnitudes, along with their errors, which are taken from Gil de Paz et al. (2003). The foreground Galactic extinctions have been corrected for these galaxies using $A_B$ values determined following Schlegel (1998) and the Galactic extinction law of Cardelli et al. (1989). The typical error in {\it B} magnitude is less than 0.05 mag, while it is around 0.1 mag in the {\it R}-band. 

For the purpose of investigating the \LZ\ relation, we first display the broad band $(B-R)$ color versus the metal abundance (derived by the $T_e$ method) and the absolute {\it B} magnitude for our BCD sample, in the left and right panels of Figure 1, respectively. In the left panel, open circles represent objects without a given measurement error of the abundance. One sees expected/familiar trends in both plots: the galaxies tend toward redder colors in progressing from lower to higher luminosity or metallicity, meanwhile with the presence of large scatters. These large scatters are mainly attributed to the internal redding and the measurement error. 

We present the $L_B$-$Z$ and $L_R$-$Z$ relations in Figures 2 and 3 respectively. Open symbols represent the objects that the measurement errors of their abundances were not given in the referenced literature.  However, the error will not affect our fitting results since we do not use it as a weighting.

In the left panel of Figure 2 we show the {\it B}-band absolute magnitude plotted versus metallicity, which was determined based on the $T_e$ method.  As we can see, a very good correlation is presented in the data.  Only two galaxies, Mrk 328 and UM 533, which have the highest metallicities in the sample, might be discrepant from the trend. In order to check whether these two galaxies are really outliers, we calculated their deviations and found that both are less than 3$\sigma$. Thus, these two points don't really deviate from the trend. The apparent discrepancy may be caused by the fact that 1) the measurement error of the metallicity for UM 533 is as large as 0.29 dex, 2) BCDs with metallicity as high as $>8.6$ are rare in our sample.  A nonweighted least-squares linear fit, using a geometrical mean functional relationship (Isobe et al. 1990), to these 66 BCDs gives
\begin{equation}
12+{\rm{log(O/H)}}_{T_e}=(4.53 \pm 0.41)-(0.218 \pm 0.038) M_B.
\end{equation} 
The fitted trend is shown as a solid line in Figure 2. The rms deviation of the data from the relationship is  0.25 dex in (O/H). The Spearman rank correlation coefficient ($\rho$, assessing how well an arbitrary monotonic function could describe the relationship between two variables) of the trend is -0.58 at a $>$4$\sigma$ level of significance, which indicates an anti-correlation between $M_B$ and O/H and means that the probability for the null hypothesis of no correlation between (O/H)$_{T_e}$ and $M_B$ is $\approx0$. The parameters of all fitting results presented in this section are summarized in Table 3.

In order to illustrate the effect of using different methods to compute the abundance on the \LZ\ relation, we plot the N2-based $L$-$Z$ relation in the right panel of the figures for each band considered. With a bit bigger scatter, a similar relationship also presents in the right panel of Figure 2.  Excluding the labelled source, Pox 4 (its deviation is larger than 3$\sigma$), we obtain the following $L_B$-$Z$ relation for the rest 73 BCDs from a geometrical mean fitting
\begin{equation}
12+{\rm{log(O/H)}_{N2}}=(4.18\pm 0.36)-(0.245\pm 0.037) M_B.
\end{equation} 
The rms deviation of the data from the relationship is 0.30 dex in (O/H).  We can see that the $T_e$-based slope is in agreement with the N2-based slope within the uncertainty.  However, as for the other two bands, the $T_e$-based slope is a bit smaller than the N2-based slope. We present a further discussion about this slope difference below.

The situation changes a bit when coming to the longer wavelength photometric band, the {\it R}-band. The trend becomes shallower and the scatter is a bit smaller. While Pox 4 is still an outlier in this band. Nonweighted least-squares linear fits to these two data sets give
\begin{equation}
12+{\rm{log(O/H)}}_{T_e}=(4.71 \pm 0.36)-(0.199 \pm 0.027) M_R.
\end{equation} 
and
\begin{equation}
12+{\rm{log(O/H)}}_{\rm{N2}}=(4.42\pm 0.32)-(0.221\pm 0.026) M_R.
\end{equation}
for the $T_e$- and N2-based $L_R$-$Z$ relations respectively.

\subsection{The \LZ\ Relation in the {\it $K_s$}-band}
The reason why we would like to arrive at an $L_{K_s}$-$Z$ relation is because such a relation should be more nearly representative of the more fundamental relationship between metal abundance and stellar mass. This is because the observed luminosity at the optical bands is dominated by recent star-formation activity rather than the stellar population that has accumulated over the galaxy's lifetime, while at the $K_s$-band the light is sensitive to the bulk of the stellar content.

The {\it $L_{K_s}$-Z} relationship displayed in Figure 4 shows a change relative to the optical ones. The slope and scatter are much smaller than those of the $L_B$- and $L_R$-$Z$ relations, which can be also seen in Table 3. This is especially true for the $T_e$-based results: The rms scatter of the {\it $L_{K_s}$-Z} relation has about one third reduction comparing with those of the optical \LZ\ relations. We obtain the following relationships using nonweighted least-squares fits to the $T_e$-based and N2-based data sets respectively,
\begin{equation}
12+{\rm{log(O/H)}}_{T_e}=(4.97 \pm 0.41)-(0.170 \pm 0.024) M_{K_s}.
\end{equation} 
and
\begin{equation}
12+{\rm{log(O/H)}_{N2}}=(4.04\pm 0.38)-(0.221\pm 0.039) M_{K_s}.
\end{equation}

\section{Discussion}
\subsection{Comparison with the Literature}
While our study clearly demonstrates a good \LZ\ relation for BCD galaxies, the $M_B-Z$ relation in Fig. 7 of HH99 is not obvious. Hopkins et al. (2002) also claimed that there is no correlation between oxygen abundance and total mass in their BCD sample. To explain the apparently contradictory results between these two studies and our work, we here compare our sample with HH99 and Hopinks' and discuss these results in the following.

1) Sample size: The sample of HH99 only contains 13 BCDs (22 for Hopkins' sample) with $\sim$4 magnitude span, while our BCD sample is $3-5$ times larger (see Table 3) and 2 magnitude wider. A small number statistics in sample size is difficult to avoid a selection bias. 

2) Uncertainties in determining O/H: HH99 used a combination of oxygen abundances determined by (four) different methods for their BCD galaxies. Despite of a larger uncertainty in empirical-based O/H comparing to $T_e$-based one, the inconsistency of the methods for determining O/H itself might cause an additional uncertainty in systematics. 

3) In Hopkins et al. (2002), although all of their metallicities are derived with $T_e$-method, they used the width of H\,{\sc i} emission-line profile at 20\% of peak, $W_{20}$, as an indicator of galaxy total mass. On the one hand, it requires radii and inclinations to calculate galaxy total masses from $W_{20}$. On the other hand, $W_{20}$ also may not be a solid indicator for rotational velocity if turbulent motions are significant, as may be expected for BCD galaxies (e.g. Thuan \& Martin 1981). 

We here used rather uniform data sets and systematically estimated O/H in two different methods, and obtained the \LZ\ relationships in three bands ($B$, $R$ \& $K_s$). Therefore, the results in the present work don't conflict with those in the literature, and BCDs do really show a well-defined \LZ\ relation.

In Table 4 we list some relevant results of the \LZ\ relations compiled from the literature, and we plot them in Fig. 5. The slope of the $L_B$-$Z$ relation for dwarf blue compact galaxies ($M_B > -18$; DBCGs; Shi et al. 2005) agrees with our results for BCDs within uncertainties. For dIs ($M_B >\,\sim -19$), the slopes of the $T_e$-based $L_B$-$Z$ relation are consistent with each other within the measurement errors (Skillman et al. 1989;  Richer \& McCall 1995; Lee et al. 2003; van Zee \& Haynes 2006), and the mean value is $-0.151$, hence much shallower than ours ($-0.218$) for BCDs. However, the difference in slope between BCDs and dIs might not be truly significant, and could be mainly due to the small number statistics in sample sizes used in the study of the \LZ\ relation. The samples of dIs always include no more than 30 (i.e. $12\sim 24$) galaxies at the {\it B}-band, and it is difficult to avoid a selection bias. It is possible that our results are less affected by sample selection given that the size of our sample is about 3 to 6 times larger.

However, diversities do appear to be significant when we compare different slopes of the $L_B$-$Z$ relation based on different empirical abundance calibrators. As shown in Table 4, all of the empirical calibrator-based slopes, except the absorption-free, KBG03-based one, for the sample ($M_B >\, \sim -22$) including massive galaxies in Salzer et al. (2005) are steeper than the $T_e$-based slopes for dIs, DBCGs and BCDs. But our N2-based slope is a bit steeper than or comparable with the absorption-free slopes from Salzer et al. (2005). At the same time, the different slopes derived with different abundance calibrators from Salzer et al. show large variations. These complicated situations might be the results of the fact that the empirical abundance calibrators are model-dependent, and thus have great complexity (see Kewley \& Ellison 2008). This reminds us that we should be cautious when comparing slopes of the \LZ\ relation derived based on different abundance calibrators. 

Similar to the optical, the $T_e$-based slope of the NIR \LZ\ relation for BCDs is consistent with that for dIs (Mendes de Oliveira et al. 2006). While for the empirical abundance calibrator-based \LZ\ relation, our N2-based slope is in agreement with the $(T_e+R_{23})$-based slope given by Vaduvescu et al. (2007). We further discuss the various possible origins resulting in the differences in the slope and scatter in the \LZ\ relation below. 

\subsection{Slope of the \LZ\ relation}
\subsubsection{The difference between $T_e$- and N2-based slopes}
As shown in section 3.2 and 3.3, the slope derived with the N2-based metallicity is {\it always} somewhat larger than that derived with the $T_e$-based metallicity for each photometric band (especially for the NIR \LZ\ relation), although they are consistent with each other within the uncertainty. In the following we discuss the possible origins of this difference.

There are two effects which can cause the slope difference. One is the uncertainty in the abundance. The empirical abundance calibrator has larger measured uncertainty relative to the direct method, and thus might result in a slope difference. This effect is important when the sample is small. The other reason is the sample itself used to study the \LZ\ relation. In our work, the N2-based sample is a bit different from the $T_e$-based sample. These two effects could simultaneously affect the \LZ\ slope.

To check this further, we use the same sample which both have direct and empirical metallicities to investigate the \LZ\ relations. In our sample, we find that there are 60 such galaxies in  the optical and 37 such galaxies in the NIR. The slopes are $-0.224\pm0.038$ and $-0.223\pm0.037$, for the $T_e$- and N2-based $L_B$-$Z$ relations respectively. These two slopes are very consistent. For the $L_R$-$Z$ relations, these two slopes are $-0.204\pm0.029$ and $-0.201\pm0.034$, which are also in accordance with each other. This indicates that almost all of the slope change between the N2- and $T_e$-based optical \LZ\ relations is due to the sample issue. 

However, it is not the case for the NIR data. For the same sample, the N2-based slope is $-0.204\pm0.039$, which is still larger than the $T_e$-based one, $-0.171\pm0.025$. But the difference becomes smaller (see section 3.3). We find that the N2 method overestimates the metallicities for several objects at the high luminosity end when we plot these two data sets in the same figure. This can be also seen from Figure 6, in which the points with circles represent galaxies with $M_{K_S} \leq -20$ mag. This overestimation (in fact,  it results in a selection effect for a small sample), combining with the sample issue, causes the large difference between the $T_e$- and N2-based slopes of the NIR relations.

\subsubsection{The NIR \LZ\ relation}
Now we return to the NIR data of BCDs. The slope of the \LZ\ relation for our BCD sample decreases monotonically as the wavelength of the photometric band used to construct it increases, which is in agreement with the result of Salzer et al. (2005). This trend can be seen directly through the green and red lines, for our and Salzer's results respectively, in Fig. 5. The dotted lines are the error-weighted linear fits to the data.  Although our slopes have larger uncertainties (due to the small size of the sample), the tendency to decrease of the \LZ\ slope with wavelength is pronounced.

The variation of the slope of the \LZ\ relation with wavelength might be expected simply due to stellar population issues. As shown by Bell \& de Jong (2001), the characteristic colors of galaxies vary smoothly with {\it M/L} ratio, in the sense that higher  {\it M/L} values are redder and lower  {\it M/L} values are bluer. Color also correlates with luminosity, as shown in Fig. 1, with bluer galaxies being less luminous and redder ones more luminous. For our metal-poor systems, as an example, a galaxy with $M_B=-12.0$ and $B-R = 0.0$ will have a $B-K$ color of $\sim$2.0, and hence  $M_K \approx -14.0$ (see Table 4 of Bell \& de Jong 2001). A galaxy with $M_B=-18.0$ and $B-R = 1.0$  will have $B-K \approx 3.0$ and $M_K \approx -21.0$. Therefore, according to the models of Bell \& de Jong (2001), one would expect a shallower slope in the {\it K}-band \LZ\ relation, since in this example $\Delta M_K=7.0$ while $\Delta M_B = 6.0$ for the same two galaxies. 

Presumably, the effect of stellar population and the lower amounts of absorption will both be acting simultaneously to reduce the observed slope in the NIR \LZ\ relation. It is not clear which effect will be the dominant one (or whether they will be of roughly equal magnitude). To check this further, we try to investigate a more fundamental relation, MZ relation. We convert luminosity into stellar mass using the $M/L$ ratio-color relation given by Bell \& de Jong (2001), which is derived by using the Bruzual \& Charlot (2003) model with a Salpter IMF (1955) and $Z=0.4 Z_\odot$. We adopt this sub-solar abundance model since our sample is metal-poor. The following three MZ relations are derived, 
\begin{equation}
\begin{array}{*{20}c}
 {12+{\rm{log(O/H)}}_{T_e}=(4.57 \pm 0.34)-(0.408 \pm 0.051) M_{\star,B},}\\
  {12+{\rm{log(O/H)}}_{T_e}=(4.57 \pm 0.34)-(0.408 \pm 0.051) M_{\star,R},}\\
   {12+{\rm{log(O/H)}}_{T_e}=(4.72 \pm 0.43)-(0.399 \pm 0.054) M_{\star,K}.}
\end{array}
\end{equation}
where $M_{\star,B}$, $M_{\star,R}$ and $M_{\star,K}$ denote that the stellar mass is obtained by using the $B$-, $R$- and $K$-band luminosity respectively. From Eq. (11) we can see that the slopes of the MZ relations are very consistent. If we use the identical sample (i.e. same as the $K$-band sample) for these three bands, the difference between these MZ slopes is even smaller, namely, $0.395\pm0.072$ for $B$- and $R$-band and $0.399\pm0.054$ for $K$-band. These results suggest that the variation of $M/L$ dominates the slope difference between the optical and NIR \LZ\ relations, while the absorption only has little effect, for our metal-poor galaxy sample. 

\subsection{Scatter in the \LZ\ Relation}
The rms scatter of the $T_e$-based \LZ\ relation is much lower for the NIR data compared to the optical data. There are $\sim 32\%$ and $\sim 26\%$ reductions in the rms scatter of the  {\it $L_{Ks}$-Z} relation relative to that of the  {\it $L_{B}$-Z} and {\it $L_{R}$-Z} relation, respectively. This indicates that the NIR data should be more suitable for studying the \LZ\ relation. However, the rms scatter of the N2-based \LZ\ relation has little change ($\leqslant10\%$). This is believed to be due to the relatively much larger uncertainties in the O/H derived with the empirical N2 method (see the following).

Since the majority of the oxygen abundances are compiled from literature, our result may suffer from the uncertainty of aperture effect. We compare metallicities derived by different authors (sources with more than one references for their metallicities in Table 2). We find that the $T_e$-based oxygen abundances determined by different works agree with each other within 0.10 dex, except for a few points with large measurement errors. However, the N2-based abundances from different works appear to have larger scatter. Therefore, aperture effect may only have a minor contribution to the scatter in the $T_e$-based \LZ\ relation on condition that we've used an error-weighted mean value. While for the N2-based \LZ\ relation, this effect should be one main source causing the scatter.

The N2-O/H relation (Eq. (4)),which itself has $\sim 0.2$ dex scatter, can also cause additional uncertainties in the N2-based \LZ\ relations. Generally, as shown in Fig. 6, the N2-based and $T_e$-based O/H is uniformly distributed around the $y=x$ line, but the standard deviation is as large as 0.19 dex. For several galaxies, the deviation can be up to 0.4-0.6 dex. Therefore, the uncertainties introduced by the N2-method itself, along with the aperture effect, should be the major reason causing the deviant points in the N2-based \LZ\ plots.

Given the fact that the typical errors are 0.05, 0.10 and 0.15 mag, in the {\it B}-, {\it R}- and $K_s$-band photometry, respectively, the observed rms scatter of 1.15, 1.12 and 1.01 mag, in the {\it B}-, {\it R}- and $K_s$-band respectively, may be affected by some other parameters. Kobulnicky et al. (2003) searched for correlations between various parameters (e.g., color, size, EW of H$\beta$, ${\rm{H}}\beta$ luminosity) and magnitude residuals ($\Delta M \equiv$\,the observed absolute magnitude $-$ the derived magnitude from the \LZ\ relation using a given metallicity) in their derived $L_B$-$Z$ relationship in order to investigate whether there are any ``second-parameter effects" that might be responsible for increased scatter in the \LZ\ relation. They found no correlations for their local sample. 

Figure 7 shows the galaxy oxygen abundance, color, EW$_{{\rm{H}}\alpha}$ and H$\alpha$ luminosity (from the left to the right) plotted versus $\Delta M$ from the best-fit linear $T_e$-based \LZ\ relation for BCDs. The bottom panels plot the full sample of 66 BCDs, while the upper two panels plot the subsample of 39 galaxies which have $K_s$ magnitude measured. Figures 7{\it a} and 7{\it i} indicate that there is a weak correlation between $\Delta M_B$ and metallicity. A Spearman rank correlation coefficient analysis gives $\rho=0.48$ and $\rho=0.41$ at a 2.9$\sigma$ and 3.3$\sigma$ level of significance, for Fig. 7{\it a} and Fig. 7{\it i} respectively. However, this correlation may be driven by the three points (in Fig. 7{\it i}) which have the highest and lowest metallicities in this sample. To test this, we do the Spearman rank correlation coefficient analysis excluding these three points and obtain $\rho=0.32$ at 2.5$\sigma$ confidence. This suggests the scatter of $L_B$-$Z$ relation does really correlate with the metallicity, although the correlation is weak.

Meanwhile, both of Figures 7{\it d} and 7{\it l} show anti-correlations between $\Delta M_B$ and $L_{{\rm{H}}\alpha}$, indicating that this correlation really exists. A Spearman rank correlation coefficient analysis gives $\rho=0.55$ at a $>4\sigma$ level of significance for Fig. 7{\it l}. We got a similar result ($\rho=0.53$ at $>4\sigma$ confidence) when excluding the two points with the largest deviations (one has a huge metallicity uncertainty and the other doesn't have a metallicity uncertainty) and the point with the largest $L_{{{\rm H}}\alpha}$. Therefore, these two correlations are not driven by some suspect points but reasonable in the sense that (1) a higher metallicity might possibly cause a larger internal extinction; (2) higher H$\alpha$ luminosity means higher blue luminosity. But the large scatters in these two correlations indicate that the observed scatter may represent some fundamental cosmic scatter (e.g. time of formation of galaxies, age dispersion, etc; see Calura et al. 2009) in the \LZ\ relation. However, no obvious correlation between the ({\it B-R}) or EW$_{{\rm{H}}\alpha}$ and $\Delta M$ can be seen in Figures 7{\it b}, 7{\it f}, 7{\it j}, 7{\it g} and 7{\it k}.  By comparing Figure 7{\it c} with Figure 7{\it k}, one can find that the correlation appearing in Figure 7{\it c} is due to the small number of statistics involved in sample.

For the $K_s$-band, there only exists one possible correlation between $\Delta M_{K_s}$and metallicity, and we can not draw a firm conclusion because of the lack of data at the low metallicity end. To check this, we compare the same data set of the {\it B}-band with the $K_s$-band, and find a similar trend for all of three samples (see Figures 7{\it a}, 7{\it e} and 7{\it i}). Therefore, \emph{the correlation between the \LZ\ relation residuals and metallicity is true, and is not due to the sample issues.} However, a Spearman rank correlation coefficient analysis indicate that the correlation between $\Delta M_{K_s}$and {\it Z} is much weaker than that in {\it B}-band. The correlation coefficient is 0.34 and only at a 2.1$\sigma$ level of significance. This is because that the NIR data suffer much less extinctions than the optical data. In contrast to the {\it B}-band, the seeming correlation between $\Delta M_{K_s}$ and $L_{{\rm{H}}\alpha}$, as shown in Figure 7{\it h}, is caused by the sample issue. This result is very reasonable since the current star forming activities should have much less effect on $L_{K_s}$ than on $L_B$.

\section{Summary}
By cross-correlating the blue compact dwarf galaxy sample of Gil de Paz et al. (2003) with the SDSS DR6, we derived and compiled oxygen abundances for a subsample containing $\sim 70$ BCD galaxies using both the direct ($T_e$) and empirical (N2) methods. The uniformity of the sample, combining with its relatively large size, makes this an excellent data set to systematically study the \LZ\ relations for BCDs for the first time. We investigate the \LZ\ relations in three different photometric bands ($B,\ R,\ {\rm{and}}\ K_s$) and explore the effects of using different abundance calibrators on the resulting \LZ\ relations. The main results are summarized bellow:

(1) We present \LZ\ relations for all three bands, from optical to near infrared, for these BCD galaxies. Our study reconciles apparently contradictory results obtained by different authors. We find that the slope of \LZ\ relation calibrated with the empirical N2-method, for each photometric band, is consistent with that evaluated using the $T_e$-based metallicities within the uncertainty. 

(2) We confirm the similar result for BCDs here that the slope of the \LZ\ relation becomes shallower with longer wavelength of the photometric band used, as has been found in larger samples that include more luminous systems (Salzer et al. 2005). 

(3) The correlation between the \LZ\ relation residuals and H$\alpha$ luminosities in the {\it B}-band, combined with the fact that no such correlation exists in the $K_s$-band, indicates that the starburst activities might be a cause of the large scatter in the $L_B$-$Z$ relation. Meanwhile, the weak correlation between the \LZ\ relation residuals and metallicity might suggest that the internal-absorption is another possible factor contributing to the scatter in the \LZ\ relationship.

\begin{acknowledgements}
The authors are very grateful to the anonymous referee for her/his careful reading and constructive comments which much improved the paper. YZHAO gratefully acknowledges financial support from the NSF of China (grant No. 10903029), the Jiangsu Planned Projects for Postdoctoral Research Funds (No. 0802031C) and the support of K.C.Wong Education Foundation, Hong Kong. Research for this project is partly supported by NSF of China (Distinguished Young Scholars No. 10425313, grants No. 10833006, \& No. 10621303), Chinese Academy of Sciences' Hundred Talent Program, and 973 project of the Ministry of Science and Technology of China (grant No. 2007CB815406). The {\it starlight} project is supported by the Brazilian agencies CNPq, CAPES and FAPESP and by the  France-Brazil CAPES/Cofecub program. This research has made use of the NASA/IPAC Extragalactic Database (NED), which is operated by the Jet Propulsion Laboratory, California Institute of Technology, under contract with the National Aeronautics and Space Administration.  This work has also made use of the VizieR catalogue access tool, CDS, Strasbourg, France.  All the authors acknowledge the work of the Sloan Digital Sky Survey (SDSS) team. Funding for the SDSS and SDSS-II has been provided by the Alfred P. Sloan Foundation, the Participating Institutions, the National Science Foundation, the U.S. Department of Energy, the National Aeronautics and Space Administration, the Japanese Monbukagakusho, the Max Planck Society, and the Higher Education Funding Council for England. The SDSS Web Site is http://www.sdss.org/. The SDSS is managed by the Astrophysical Research Consortium for the Participating Institutions. The Participating Institutions are the American Museum of Natural History, Astrophysical Institute Potsdam, University of Basel, University of Cambridge, Case Western Reserve University, University of Chicago, Drexel University, Fermilab, the Institute for Advanced Study, the Japan Participation Group, Johns Hopkins University, the Joint Institute for Nuclear Astrophysics, the Kavli Institute for Particle Astrophysics and Cosmology, the Korean Scientist Group, the Chinese Academy of Sciences (LAMOST), Los Alamos National Laboratory, the Max-Planck-Institute for Astronomy (MPIA), the Max-Planck-Institute for Astrophysics (MPA), New Mexico State University, Ohio State University, University of Pittsburgh, University of Portsmouth, Princeton University, the United States Naval Observatory, and the University of Washington.
\end{acknowledgements}

\newpage
\clearpage
\begin{deluxetable}{lccccc}
\tablenum{1}
\tablecaption{Electron temperatures, number densities and oxygen abundances of BCDs in SDSS}
\tablewidth{0pc}
\tabletypesize{\scriptsize}
\tablehead{
\colhead{}&\colhead{$T_e$}&\colhead{$n_e$}&\multicolumn{3}{c}{12+log(O/H)}\\
\cline{4-6}
\colhead{Object Name}&\colhead{(K)}&\colhead{(cm$^{-3}$)}&\colhead{($T_e$)}&\colhead{(N2)}&\colhead{(Other; $T_e$)}
}
\startdata
HS 0822+3542\dotfill   &18665.&325.&   7.41&7.49&7.42\tablenotemark{b}\\
HS 1400+3927\dotfill &12335.&88.  &8.11 &  8.06&\\
HS 1440+4302\dotfill & 12544.&49.& 8.10 &8.05&8.07\tablenotemark{c}\\
HS 1609+4827\dotfill &\nodata&\nodata&\nodata&8.37&\\
Haro 3\dotfill & 9341.&57.& 8.56   &8.46&8.29\tablenotemark{b}, 8.37\tablenotemark{c}, 8.37\tablenotemark{d}\\
II Zw 71\dotfill&\nodata&\nodata&\nodata&8.40&\\
Mrk 1313\dotfill & 11921.&  26.&8.17  & 8.08 &8.19\tablenotemark{b}, 8.34\tablenotemark{e}\\
Mrk 1416\dotfill   & 14453.&100.\tablenotemark{a}& 7.82 &7.99&7.85\tablenotemark{c}\\
Mrk 1418\dotfill &\nodata&\nodata&\nodata&8.48&\\
Mrk 1423\dotfill &\nodata&\nodata&\nodata&8.52&\\
Mrk 1480\dotfill &\nodata&\nodata&\nodata&8.03&\\
Mrk 1481\dotfill &\nodata&\nodata&\nodata&8.11&\\
Mrk 178\dotfill &\nodata&\nodata&\nodata&7.86&\\
Mrk 409\dotfill &\nodata&\nodata&\nodata&8.81&\\
SBS 1054+504\dotfill  & 10610.&100.\tablenotemark{a}&8.26&8.34& \\
SBS 1147+520\dotfill &\nodata&\nodata&\nodata&8.19&\\
SBS 1428+457\dotfill &9697.&95.& 8.44  &8.18&8.40\tablenotemark{b}\\
SBS 1533+574\dotfill  &11690.&21.& 8.14 &8.13&8.00\tablenotemark{c}\\
UGCA 184\dotfill    &12825.&100.\tablenotemark{a}&   8.04&7.86&8.04\tablenotemark{b}, 8.00\tablenotemark{b}\\
UM 439\dotfill     & 12099.  &24.&8.09 &8.06& 8.07\tablenotemark{c}, 8.05\tablenotemark{f}\\
UM 452\dotfill     &  10306. &100.\tablenotemark{a} &8.27    &8.40&\\
UM 456A\dotfill &\nodata&\nodata&\nodata&8.28&\\
UM 491\dotfill &\nodata&\nodata&\nodata&8.31&\\
VCC 0130\dotfill &\nodata&\nodata&\nodata&8.40&\\
\enddata
\tablenotetext{a}{$n_e$=100. cm$^{-3}$ is an assumed nominal value.}
\tablerefs{
$^{\rm{b}}$ Izotov et al. (2006); $^{\rm{c}}$ Nagao  et al. (2006); $^{\rm{d}}$ Steel et al. (1996); $^{\rm{e}}$ 
Masegosa et al. (1994); $^{\rm{f}}$  Kobulnicky \& Skillman (1996).
}

\end{deluxetable}

\newpage
\clearpage
\begin{deluxetable}{lcc ccc ccc r}
\tablenum{2}
\tabletypesize{\scriptsize}
\tablecaption{Sample and properties of BCDs used to study the \LZ\ relationships}
\tablewidth{0pt}
\tablehead{ 
\colhead{}&\colhead{DM\tablenotemark{a}}&\colhead{{\it B}\tablenotemark{a}}&\colhead{ {\it R}\tablenotemark{a}}& \colhead{$K_s$\tablenotemark{b}}&\multicolumn{2}{c}{12+log(O/H)}&\colhead{}&\colhead{}&\colhead{}\\
\cline{6-7}
\colhead{Object Name} &\colhead{ (mag)} & \colhead{(mag)} & \colhead{(mag)} & \colhead{(mag)} & \colhead{$T_e$}& \colhead{N2}&\multicolumn{3}{c}{Reference}\\
\cline{8-10}
\colhead{(1)}&\colhead{(2)}&\colhead{(3)}&\colhead{(4)}&\colhead{(5)}&\colhead{(6)}&\colhead{(7)}&\colhead{(8)}&\colhead{(9)}&\colhead{(10)}
}
\startdata
 HS 0029+1748\dotfill&32.61& $16.74\pm 0.05$ & $15.93\pm 0.14$ & \nodata & $ 8.05\pm 0.02$ &  7.87  &\nodata&5&28\\
 HS 0822+3542\dotfill&30.03& $17.85\pm 0.03$ & $17.76\pm 0.14$ & \nodata & $ 7.42\pm 0.03$ &   7.47 &\nodata&6, 7 &6, 7, 29\\
 HS 1400+3927\dotfill&31.62& $17.04\pm 0.05$ & $15.94\pm 0.10$ & \nodata & $ 8.11$ &  8.08  &\nodata& 6 &6\\
 HS 1440+4302\dotfill&32.90& $17.58\pm 0.05$ & $16.54\pm 0.11$ & \nodata & $ 8.09\pm 0.03$ & 8.08  &\nodata	&5, 6 &6, 29, 30\\
 HS 1442+4250\dotfill&30.45& $15.55\pm 0.05$ & $15.01\pm 0.07$ & \nodata & $ 7.61\pm 0.02$ & 7.60  &\nodata	&  5& 31\\
 HS 1609+4827\dotfill&33.15& $15.11\pm 0.11$ & $14.55\pm 0.09$ & $13.15\pm 0.15$ & $ 8.14\pm 0.14$ & 8.37  &  1	&7&6, 7\\
 Haro 14\dotfill & 30.73 &$13.65\pm 0.05$ &$12.91\pm 0.14$  &10.95& \nodata& 8.38  & 2 & \nodata &29\\
       Haro 2\dotfill&31.67& $13.39\pm 0.04$ & $12.87\pm 0.11$ & $10.43\pm 0.04$ & $ 8.38$ &  8.64  & 2&  8, 9&29\\
       Haro 3\dotfill&30.79& $13.22\pm 0.04$ & $12.61\pm 0.13$ & $10.61\pm 0.05$ & $ 8.46\pm 0.10$ & 8.39&   &5, 6, 7, 10&6, 28, 29\\
       Haro 4\dotfill&29.73& $15.59\pm 0.04$ & $15.44\pm 0.17$ & \nodata & $ 7.82\pm 0.02$ & 7.85	&\nodata &  5&29\\
       Haro 8\dotfill&31.02& $14.27\pm 0.03$ & $13.20\pm 0.11$ & $12.02\pm 0.14$ & $ 8.35\pm 0.04$ & 8.16& 1 &7&7\\
       Haro 9\dotfill&30.93& $13.02\pm 0.05$ & $12.24\pm 0.10$ & $10.41\pm 0.04$ & $ 8.40$ & 8.60& 1&   9&29\\
      I Zw 123\dotfill&30.34& $15.42\pm 0.10$ & $14.85\pm 0.08$ & $13.22\pm 0.14$ & $ 8.07\pm 0.02$ & 7.99&  1 &5, 7, 11&7, 29, 32\\
       I Zw 18\dotfill&30.50& $16.05\pm 0.04$ & $16.24\pm 0.07$ & \nodata & $ 7.18\pm 0.01$ & 7.61	&\nodata   &   5, 12&29, 32\\
      II Zw 40\dotfill&29.96& $11.87\pm 0.04$ & $11.10\pm 0.09$ & $12.35\pm 0.11$ & $ 8.09\pm 0.01$ & 7.94& 1  &12, 13&29\\
      II Zw 70\dotfill&31.36& $14.84\pm 0.12$ & $14.29\pm 0.13$ & $12.80\pm 0.12$ & $ 8.18\pm 0.08$ & 8.18&  1 &12, 14, 15&14, 33\\
      II Zw 71\dotfill&31.36& $14.45\pm 0.15$ & $13.54\pm 0.12$ & $12.05\pm 0.10$ & $ 8.24$ & 8.42& 1  &  8&6, 29\\
     Mrk 0005\dotfill&30.60& $15.13\pm 0.04$ & $14.56\pm 0.13$ & $12.68$ & $ 8.06\pm 0.04$ & 8.19	& 3  &  5 &29, 34\\
      Mrk 108\dotfill&31.69& $15.15\pm 0.03$ & $14.66\pm 0.08$ & $13.76\pm 0.20$ & $ 7.96\pm 0.02$ & 8.03& 1  &15, 16&29\\
     Mrk 1313\dotfill&32.51& $16.02\pm 0.03$ & $15.50\pm 0.06$ & $13.63\pm 0.27$ & $ 8.22\pm 0.10$ & 8.05&1   & 6, 7, 18&6, 7, 29\\
     Mrk 1329\dotfill&31.02& $14.08\pm 0.03$ & $13.38\pm 0.09$ & $11.93$ & $ 8.26\pm 0.02$ & 7.95	&2   &  5,  13&28, 29\\
     Mrk 1416\dotfill&32.64& $16.32\pm 0.03$ & $15.75\pm 0.02$ & \nodata & $ 7.84\pm 0.02$ & 7.94&\nodata   &5, 6&6, 32\\
     Mrk 1418\dotfill& 30.28  &$13.86\pm 0.03$&$ 12.61\pm 0.19$  &$11.604\pm 0.061$ &\nodata& 8.48 &1 & \nodata&6\\
Mrk 1423\dotfill&31.55& $   14.90\pm 0.03$&$ 13.52\pm 0.10$& $12.65\pm 0.11$&\nodata&  8.52 &1 &\nodata &6\\
     Mrk 1434\dotfill&32.62& $16.77\pm 0.04$ & $16.43\pm 0.13$ & \nodata & $ 7.79\pm 0.01$ & 7.69&\nodata   &5&29\\
     Mrk 1450\dotfill&30.83& $15.75\pm 0.05$ & $15.09\pm 0.09$ & \nodata & $ 7.96\pm 0.02$ & 7.93& \nodata  &5&35\\
     Mrk 1480\dotfill&32.18& $16.17\pm 0.03$ & $15.56\pm 0.08$ & $13.66\pm 0.18$ & $ 8.04\pm 0.05$ & 8.03& 1  &7&6, 7\\
     Mrk 1481\dotfill &32.18&$ 16.19\pm 0.03$&$ 15.57\pm 0.06$ &\nodata&\nodata& 8.11 &\nodata &\nodata &6\\
      Mrk 178\dotfill&28.11& $14.15\pm 0.04$ & $13.60\pm 0.09$ & $11.74$ & $ 7.92\pm 0.02$ &7.88 & 4  &5, 12&6, 29\\
      Mrk 209\dotfill&28.82& $14.15\pm 0.03$ & $13.94\pm 0.13$ & $12.70$ & $ 7.76\pm 0.01$ &7.66&  4 &  5 &29\\
      Mrk 324\dotfill&32.01& $15.17\pm0.03$&$ 14.60\pm 0.13$ &$12.50\pm  0.06 $ &\nodata& 8.18& 3& \nodata&29\\
    Mrk 328\dotfill&31.75& $14.93\pm 0.05$ & $14.18\pm 0.09$ & $12.10\pm 0.08$ & $ 8.66$ & 8.64& 1  & 8&29\\
    Mrk 409\dotfill&31.64&$14.37\pm 0.03 $&$13.33\pm 0.05$& $11.17\pm 0.03$&\nodata&  8.80 &1 &\nodata &6\\
      Mrk 450\dotfill&30.57& $14.44\pm 0.05$ & $13.65\pm 0.10$ & $11.97$ & $ 8.12\pm 0.02$ & 8.09& 3  &   5 &28, 29\\
      Mrk 475\dotfill&29.93& $16.20\pm 0.03$ & $15.67\pm 0.07$ & \nodata & $ 7.93\pm 0.02$ & 7.89	&\nodata   &  5 &35 \\
      Mrk 600\dotfill&30.82& $14.85\pm 0.03$ & $14.82\pm 0.15$ & $13.11$ & $ 7.94\pm 0.06$ & 7.84& 3  &  5, 12, 17 &29, 34\\
      Mrk 67\dotfill&30.77& $16.10\pm 0.03$ & $15.34\pm 0.08$ & $14.38\pm 0.26$ & $ 8.08\pm 0.08$ & 7.90& 1 &  5, 11&28, 29\\
      Mrk 709\dotfill&30.98& $16.32\pm 0.03$ & $15.65\pm 0.02$ & $14.14\pm 0.27$ & $ 7.68\pm 0.04$ &8.38& 1  &18& 29\\
      Mrk 86\dotfill&29.20&$12.07\pm 0.03$&$ 11.49\pm 0.11$&$9.13 \pm 0.03$&\nodata&  8.53  &1&\nodata&29\\
      Mrk 900\dotfill&31.37& $14.17\pm 0.03$ & $13.56\pm 0.12$ & $11.38\pm 0.10$ & $ 8.07\pm 0.03$ & 8.48&  1 &   19&29\\
      Mrk 996\dotfill&31.88 &$15.01\pm0.03$&$14.08\pm 0.14$&11.85 &\nodata&8.21  &3 &\nodata &29\\
     NGC 1522\dotfill&30.13& $14.03\pm 0.01$ & $13.22\pm 0.06$ & $11.38\pm 0.10$ & $ 8.07\pm 0.05$ &\nodata& 1  &  18&
     \nodata\\
     NGC 1705\dotfill&28.54& $13.09\pm 0.01$ & $12.19\pm 0.06$ & $10.52\pm 0.06$ & $ 8.21\pm 0.05$ &\nodata& 1  &  20&\nodata\\
     NGC 2915\dotfill&27.78&$11.93\pm 0.01$&$10.96\pm 0.06$&$9.83\pm 0.06$&\nodata& 8.40 & 1&\nodata &  36\\
     NGC 3125\dotfill&29.84& $13.05\pm 0.01$ & $12.25\pm 0.06$ & $10.52\pm 0.05$ & $ 8.29\pm 0.10$ &8.01& 1  & 9, 12, 18, 21&29\\
     NGC 4861\dotfill&30.50& $12.68\pm 0.03$ & $11.91\pm 0.09$ & $11.77\pm 0.11$ & $ 8.00\pm 0.01$ &7.90&  1 & 5, 12, 22&29, 32\\
      Pox 186\dotfill&30.93& $17.73\pm 0.03$ & $17.39\pm 0.01$ & \nodata & $ 7.74\pm 0.01$ &\nodata &\nodata&   12, 23&\nodata\\
 Pox 4\dotfill&33.45& $15.27\pm 0.04$ & $14.88\pm 0.10$ & $13.52$ & $ 7.98\pm 0.06$ &7.67&  2 &   12, 21, 23&29\\
SBS 0940+544C\dotfill&31.91& $17.18\pm 0.04$ & $17.05\pm 0.10$ & \nodata & $ 7.43\pm 0.02$ &7.50&  \nodata &   5&29\\
 SBS 1054+504\dotfill&31.52& $16.08\pm 0.04$ & $15.46\pm 0.15$ & $13.13\pm 0.14$ & $ 8.26$ &8.34&1   & 6&6\\
 SBS 1147+520\dotfill&31.39&$  16.95\pm 0.05$&$ 15.98\pm 0.09$&\nodata&\nodata& 8.19 &\nodata & \nodata&6\\
 SBS 1331+493\dotfill&29.98& $14.87\pm 0.03$ & $14.16\pm 0.17$ & \nodata & $ 7.78\pm 0.02$ &7.89&\nodata   &   5&29, 35\\
 SBS 1415+437\dotfill&30.02& $15.43\pm 0.03$ & $14.77\pm 0.12$ & \nodata & $ 7.59\pm 0.01$ &7.76& \nodata  &  5, 24 &29, 34\\
 SBS 1428+457\dotfill&32.74& $15.42\pm 0.05$ & $14.67\pm 0.09$ & $13.15\pm 0.14$ & $ 8.42\pm 0.05$ &8.18& 1  &  6, 7&6\\
 SBS 1533+574\dotfill&33.48& $16.02\pm 0.10$ & $15.30\pm 0.10$ & $13.74\pm 0.21$ & $ 8.07\pm 0.07$ & 8.11&  1 &  5, 6&6, 32\\
      Tol 002\dotfill&29.25& $14.06\pm 0.01$ & $13.34\pm 0.06$ & $12.47\pm 0.13$ & $ 7.98\pm 0.02$ & 8.15&  1 &  12, 18&29\\
 Tol 1345-420\dotfill&32.54& $15.87\pm 0.03$ & $15.00\pm 0.03$ & $14.11\pm 0.22$ & $ 7.99\pm 0.02$ &7.85& 1  &  12, 18&29\\
 Tol 1434+032\dotfill&31.66& $15.86\pm 0.03$ & $15.42\pm 0.06$ & \nodata & $ 7.88\pm 0.04$ &7.87&\nodata   &  7 &7\\
       Tol 17\dotfill&32.19& $15.99\pm 0.05$ & $15.05\pm 0.06$ & \nodata & $ 7.94\pm 0.05$ &8.17& \nodata  &  18&29\\
       Tol 35\dotfill&32.16& $14.17\pm 0.03$ & $13.22\pm 0.05$ & $11.73\pm 0.08$ & $ 8.19\pm 0.01$ &8.08& 1  & 12, 18&29\\
       Tol 65\dotfill&32.78& $17.26\pm 0.04$ & $16.84\pm 0.03$ & $15.99$ & $ 7.50\pm 0.01$ & 7.51& 2  & 5, 12, 18 &29, 37\\
       Tol 85\dotfill&33.47& $16.51\pm 0.03$ & $15.99\pm 0.01$ & \nodata & $ 8.04\pm 0.01$ & \nodata& \nodata & 18&\nodata\\
UCM 1612+1308\dotfill&33.63& $17.21\pm 0.14$ & $17.31\pm 0.07$ & \nodata & $ 8.17\pm 0.03$ &7.96& \nodata  &  25&29\\
     UGCA 184\dotfill&31.81& $15.99\pm 0.04$ & $15.80\pm 0.10$ & $14.17\pm 0.22$ & $ 8.03\pm 0.01$ &7.89& 1  &  5, 6, 7&6, 35\\
     UGCA 412\dotfill&33.07& $15.55\pm 0.05$ & $14.62\pm 0.09$ & \nodata & $ 8.12$ &8.38&\nodata   &  8&8\\
       UM 133\dotfill&31.91& $15.41\pm 0.03$ & $14.51\pm 0.14$ & \nodata & $ 7.69\pm 0.02$ & 7.78& \nodata  &   5&28\\
       UM 323\dotfill&32.26& $16.09\pm 0.04$ & $15.24\pm 0.08$ & \nodata & $ 7.96\pm 0.04$ &8.00& \nodata  &   7&7\\
       UM 382\dotfill&33.53& $18.20\pm 0.04$ & $17.83\pm 0.22$ & \nodata & $ 7.82\pm 0.03$ &\nodata& \nodata  & 26&\nodata\\
       UM 408\dotfill&33.58& $17.46\pm 0.16$ & $16.70\pm 0.12$ & \nodata & $ 7.74\pm 0.05$ & 7.90& \nodata  &   18,  27 &27\\
       UM 439\dotfill&30.73& $14.77\pm 0.03$ & $14.09\pm 0.06$ & \nodata & $ 8.08\pm 0.03$ &7.93&  \nodata &   5, 6, 12&6, 28\\
       UM 452\dotfill&31.45& $15.25\pm 0.03$ & $14.07\pm 0.11$ & $12.92\pm 0.17$ & $ 8.27$ & 8.40& 1  & 6&6, 29\\
       UM 455\dotfill&33.64& $17.02\pm 0.03$ & $16.26\pm 0.01$ & \nodata & $ 7.74\pm 0.02$ &\nodata& \nodata  &   18&\nodata\\
       UM 456A\dotfill&31.95&$16.71\pm 0.03$&$ 16.05\pm0.01$&\nodata&\nodata& 8.21  &\nodata &\nodata &6\\
UM 491\dotfill&32.20&$15.54\pm 0.03$&$ 14.95\pm0.07$& $13.43\pm 0.21$&\nodata&  8.31 &1 &\nodata &6 \\
       UM 533\dotfill&30.34& $14.63\pm 0.04$ & $13.64\pm 0.11$ & \nodata & $ 8.55\pm 0.29$ &8.32& \nodata  &  7&7\\
       VCC 0130\dotfill&31.02 &$17.05\pm 0.05$&$ 16.27\pm 0.01$&\nodata&\nodata&8.28 &\nodata &\nodata &6 \\
     VCC 0459\dotfill&31.02& $14.95\pm 0.05$ & $14.13\pm 0.01$ & $12.45$ & $ 8.27\pm 0.09$ & 8.28& 4  & 4& 4\\
     VCC 0655\dotfill&31.02&$13.32\pm 0.04$&$ 12.12\pm 0.03$&$10.48\pm 0.02$&\nodata& 8.72 & 1&\nodata &38\\
     VCC 0848\dotfill&31.02& $15.03\pm 0.05$ & $14.10\pm 0.01$ & $12.71$ & $ 8.06\pm 0.12$ & 8.27&  4&  5& 38\\
    VII Zw 403\dotfill&28.41& $14.11\pm 0.04$ & $13.58\pm 0.11$ & $12.48$ & $ 7.70\pm 0.01$ & 7.84&  4 &  5&32\\
\enddata
\tablecomments{
Col. (1): Galaxy name.  Col. (2): Distance moduli. Col. (3) -- (5): {\it B, R} and {\it $K_s$}-band apparent magnitudes, respectively. Col. (6): Oxygen abundances determined by the $T_e$ method. Col (7): Oxygen abundances determined by the N2 method. Col. (8): References for the $K_s$ magnitudes. Col. (9): References for the metal abundances derived by the $T_e$ method. Col. (10): References for the N2 ratios.}
\tablenotetext{a}{All data are from Gil de Paz et al. (2003), and all of the magnitudes have been corrected for the 
Galactic extinctions.}
\tablenotetext{b}{The Galactic extinctions are not corrected for the magnitudes from 2MASS, while the corrections have been applied for the magnitudes from the other three sources.}
\tablerefs{
(1) Jarrett et al. 2000; (2) Noeske et al. 2003; (3) Noeske et al. 2005; (4) Vaduvescu et al. 2007; (5) Nagao  et al. 2006; (6) This work; (7) Izotov et al. 2006; (8) Shi et al. 2005; (9) Mas-Hesse \& Kunth 1999; (10) Steel et al. 1996; (11) Garnett 1990; (12) Kobulnicky \& Skillman 1996; (13) Guseva et al. 2000;  (14) French 1980;  (15) Alloin et al. 1979; (16) van Zee et al. 1998; (17) Augarde et al. 1990; (18) Masegosa et al. 1994; (19) van Zee \& Haynes 2006; (20) Lee \& Skillman 2004; (21) Vacca \& Conti 1992; (22) Koubulnicky \& Skillman 1998; (23) Guseva et al. 2007; (24) Thuan et al. 1999; (25) Rego et al. 1998; (26) Kniazev et al. 2001; (27) Pustilnik et al. 2002; (28) Izotov \& Thuan 2004; (29) Gil de Paz et al. 2003; (30) Melbourne et al. 2004; (31) Guseva et al. 2003; (32) Izotov et al. 1997; (33) Jansen et al. 2000; (34) Izotov et al. 1998; (35) Izotov et al. 1994; (36) Lee et al. 2003; (37) Izotov et al. 2004; (38)  V\'ichez \& Iglesias-P\'aramo 2003.}
\end{deluxetable}

\tablewidth{0pc}
\begin{deluxetable}{cccccccc}
\tablecaption{Results of linear fits to the \LZ\ relations}
\tablenum{3}
\tablehead{
\colhead{Method}&\colhead{Band}&\colhead{No. of Galaxies}&\colhead{Intercept} &\colhead{Slope} &\colhead{ rms}&\colhead{$\rho$}&\colhead{Significance}\\
\colhead{(1)}&\colhead{(2)}&\colhead{(3)}&\colhead{(4)}&\colhead{(5)}&\colhead{(6)}&\colhead{(7)}&\colhead{(8)}}
\startdata
\multirow{3}{*}{$T_e$}&{\it B}&66&$4.53\pm0.41$&$-0.218\pm0.038$&0.25&-0.58&$>$4$\sigma$\\
&{\it R}&66&$4.71 \pm 0.36$&$-0.199 \pm 0.027$&0.23&-0.63&$>$4$\sigma$\\
&{\it $K_s$}&39&$4.97 \pm 0.41$&$-0.170 \pm 0.024$&0.17&-0.68&$>$4$\sigma$\\
\hline
\multirow{3}{*}{N2}&{\it B}&73&$4.18 \pm 0.36$&$-0.245 \pm 0.037$&0.30&-0.53&$>$4$\sigma$\\
&{\it R}&73&$ 4.42 \pm 0.32$&$ -0.221 \pm 0.026$&0.29&-0.60&$>$4$\sigma$\\
&{\it $K_s$}&47&$ 4.04 \pm 0.38$&$ -0.221 \pm 0.039$&0.27&-0.56&$$3.7$\sigma$\\
\enddata
\tablecomments{
Col. (1): Method for determining the oxygen abundance. Col (2): Wavelength band. Col (3): Number of galaxies used in the fit. Col (4): \LZ\ relation intercept. Col (5): \LZ\ relation slope. Col. (6): rms error in log(O/H). Col. (7):  Spearman rank correlation coefficient. Col (8): Level of significance for the Spearman rank correlation coefficient.}
\end{deluxetable}
\clearpage

\tablewidth{0pt}
\begin{deluxetable}{lccc cc}
\tablecaption{Luminosity-metallicity relations for star-forming galaxies}
\tablenum{4}
\tabletypesize{\scriptsize}
\tablehead{
\colhead{Sample}&\colhead{Method}&\colhead{Intercept} &\colhead{Slope} &\colhead{ rms}&\colhead{Reference}\\
\colhead{(1)}&\colhead{(2)}&\colhead{(3)}&\colhead{(4)} &\colhead{(5)} &\colhead{(6)}}
\startdata
\multicolumn{6}{c}{{\it B}-band}\\
\hline
19 dIs\dotfill&$T_e$&5.50&$-0.153$&0.16&SKH89\\
12 dIs\dotfill&$T_e$&$5.67\pm0.48$&$-0.147\pm0.029$&0.09&RM95\\
22 dIs\dotfill&$T_e$&$5.59\pm0.54$&$-0.153\pm0.025$&0.18&LMK03\\
24 BCGs $M_B > -18$\dotfill&$T_e$&$5.50\pm0.81$&$-0.16\pm0.05$&$> 0.2$&SKL05\\
21 dIs\dotfill&$T_e$&$5.65\pm0.17$&$-0.149\pm0.011$&0.15&ZH06\\
66 BCDs\dotfill&$T_e$&$4.53\pm0.41$&$-0.218\pm0.038$&0.25&Present work\\
29 dIs\dotfill&$T_e+P$&$5.80\pm0.17$&$-0.139\pm0.011$&$\sim0.15$&PVC04\\
24 BCGs $M_B > -18$\dotfill&N2&$5.16\pm0.63$&$-0.19\pm0.05$&$> 0.2$&SKL05\\
73 BCDs\dotfill&N2&$4.18 \pm 0.36$&$-0.245 \pm 0.037$&0.30&Present work\\
54 Spirals\dotfill&$P$&$6.93\pm0.37$&$-0.079\pm0.018$&$\sim0.2$&PVC04\\
$\sim 53000$ SFGs\dotfill&SDSS $R_{23}$&$5.238\pm0.018$&$-0.185\pm0.001$&$> 0.3$&THK04\\
48 BCGs $M_B < -18$\dotfill&$T_e$&$7.37\pm0.53$&$-0.05\pm0.03$&$> 0.2$&SKL05\\
48 BCGs $M_B < -18$\dotfill&N2&$7.06\pm0.63$&$-0.08\pm0.03$&$> 0.2$&SKL05\\
765 ELGs\dotfill&EP84 $R_{23}$&$3.75\pm0.05$&$-0.280\pm0.003$&0.29&SLM05\\
765 ELGs\dotfill&KBG03 $R_{23}$&$4.56\pm0.05$&$-0.222\pm0.003$&0.25&SLM05\\
765 ELGs\dotfill&SDSS $R_{23}$&$3.96\pm0.05$&$-0.271\pm0.003$&0.28&SLM05\\
765 ELGs\dotfill&EP84 $R_{23}$&$3.75\pm0.05$&$-0.253\pm0.002$&0.26&SLM05\tablenotemark{a}\\
765 ELGs\dotfill&KBG03 $R_{23}$&$4.56\pm0.05$&$-0.200\pm0.002$&0.22&SLM05\tablenotemark{a}\\
765 ELGs\dotfill&SDSS $R_{23}$&$3.96\pm0.05$&$-0.247\pm0.002$&0.25&SLM05\tablenotemark{a}\\
\hline
\multicolumn{6}{c}{{\it R}-band}\\
\hline
765 ELGs\dotfill&EP84 $R_{23}$&$3.42\pm0.05$&$-0.266\pm0.002$&0.27&SLM05\tablenotemark{b}\\
765 ELGs\dotfill&KBG03 $R_{23}$&$4.28\pm0.06$&$-0.211\pm0.003$&0.23&SLM05\tablenotemark{b}\\
765 ELGs\dotfill&SDSS $R_{23}$&$3.64\pm0.05$&$-0.259\pm0.002$&0.26&SLM05\tablenotemark{b}\\
66 BCDs\dotfill&$T_e$&$4.71 \pm 0.36$&$-0.199 \pm 0.027$&0.23&Present work\\
73 BCDs\dotfill&N2&$ 4.42 \pm 0.32$&$ -0.221 \pm 0.026$&0.29&Present work\\
\hline
\multicolumn{6}{c}{$K_s$-band}\\
\hline
370 ELGs\dotfill&EP84 $R_{23}$&$3.92\pm0.09$&$-0.212\pm0.003$&0.24&SLM05\\
370 ELGs\dotfill&KBG03 $R_{23}$&$4.03\pm0.10$&$-0.195\pm0.004$&0.23&SLM05\\
370 ELGs\dotfill&SDSS $R_{23}$&$4.85\pm0.09$&$-0.173\pm0.003$&0.18&SLM05\\
27 dIs\dotfill&$T_e$&$5.37\pm0.20$&$-0.151\pm0.013$&0.14&OTC06\tablenotemark{c}\\
25 dIs\dotfill&$T_e+R_{23}$&$5.58\pm0.24$&$-0.141\pm0.015$&0.10&VMR07\\
14 BCDs\dotfill&$Te+R_{23}$&$4.21\pm0.54$&$-0.224\pm0.030$&0.11&VMR07\tablenotemark{d}\\
39 BCDs\dotfill&$T_e$&$4.97 \pm 0.41$&$-0.170 \pm 0.024$&0.17&Present work\\
47 BCDs\dotfill&N2&$4.04 \pm 0.38$&$ -0.221 \pm 0.039$&0.27&Present work\\
\enddata
\tablecomments{
Col. (1): Sample: dIs, dwarf irregular galaxies; SFGs, star-forming galaxies;
BCGs, blue compact galaxies; ELGs, emission line galaxies. Col. (2): Method for determining the oxygen abundance. Col (3): \LZ\ relation intercept. Col (4): \LZ\ relation slope. Col. (5): rms error in log(O/H). Col. (6): Reference. }
\tablenotetext{a}{These results are corrected for the internal absorption in luminosity.}
\tablenotetext{b}{Results of the \LZ\ relation for $V$-band.}
\tablenotetext{c}{Results re-estimated by us using a geometrical mean fit to their 27 data points (excluding the two points with metallicity derived by the $R_{23}$ method).}
\tablenotetext{d}{It is worth noting that the definition of BCDs used in their work is different from that in the present work.}
\tablerefs{
(SKH89) Skillman et al. 1989; (RM95) Richer \& McCall 1995; (LMK03) Lee et al. 2003; (SKL05) Shi et al. 2005; (ZH06) van Zee \& Haynes 2006; (PVC04) Pilyugin et al. 2004; (THK04) Tremonti et al. 2004; (SLM05) Salzer et al. 2005; (OTC06) Mendes de Oliveira et al. 2006; (VMR07) Vaduvescu et al. 2007.}
\end{deluxetable}

\begin{figure}
\centering
\includegraphics[width=1.\textwidth]{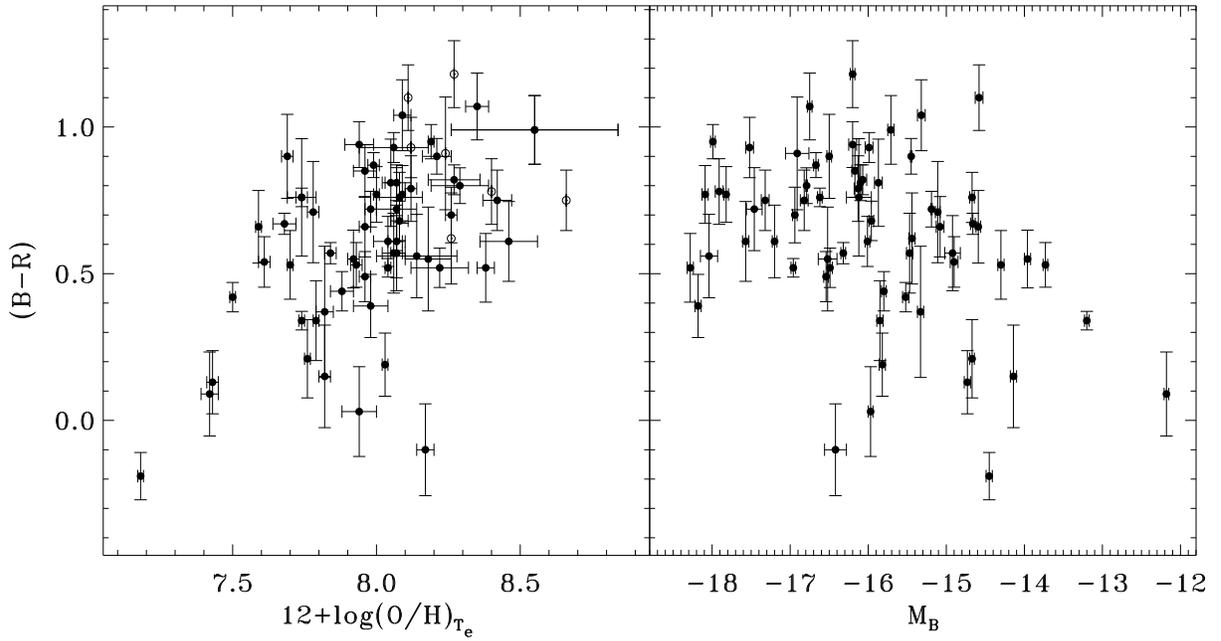}
\caption{{\it Left panel}: The ($B-R$) color vs. metallicity diagram. Open circles show objects with errors of the abundances unavailable. {\it Right panel}: The ($B-R$) color vs. the {\it B}-band absolute magnitude of the BCD galaxies. The familiar trends (i.e. objects with higher luminosity or metallicity having redder color) are clearly present in these two plots. }
\label{Fig1}
\end{figure}

\newpage 
\clearpage
\begin{figure}
\centering
\includegraphics[width=1.\textwidth]{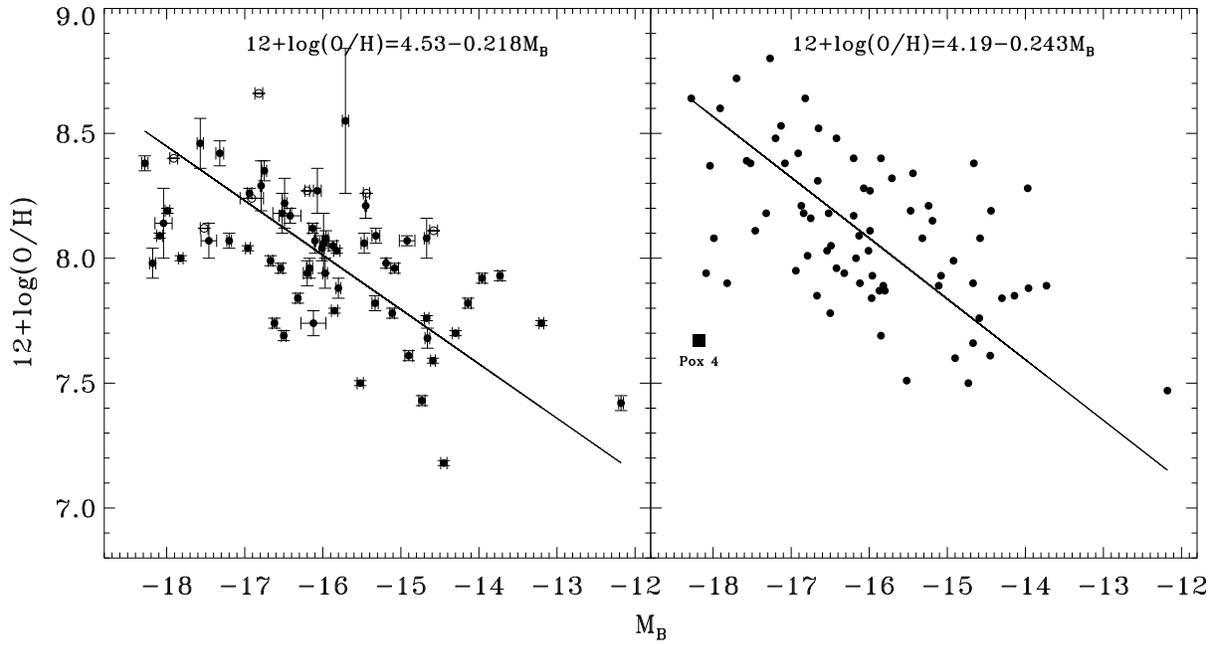}
\caption{{\it B}-band \LZ\ relationships for the BCDs. The objects without measurement errors in abundances and/or magnitudes are plotted with open symbols. Larger symbols show objects which are not used to construct the \LZ\ relations. The solid lines represent geometrical mean, least-squares linear fits to the data. {\it Left panel}: $T_e$-based results. {\it Right panel}: N2-based results.}
\label{Fig2}
\end{figure}

\newpage 
\clearpage
\begin{figure}
\centering
\includegraphics[width=1.\textwidth]{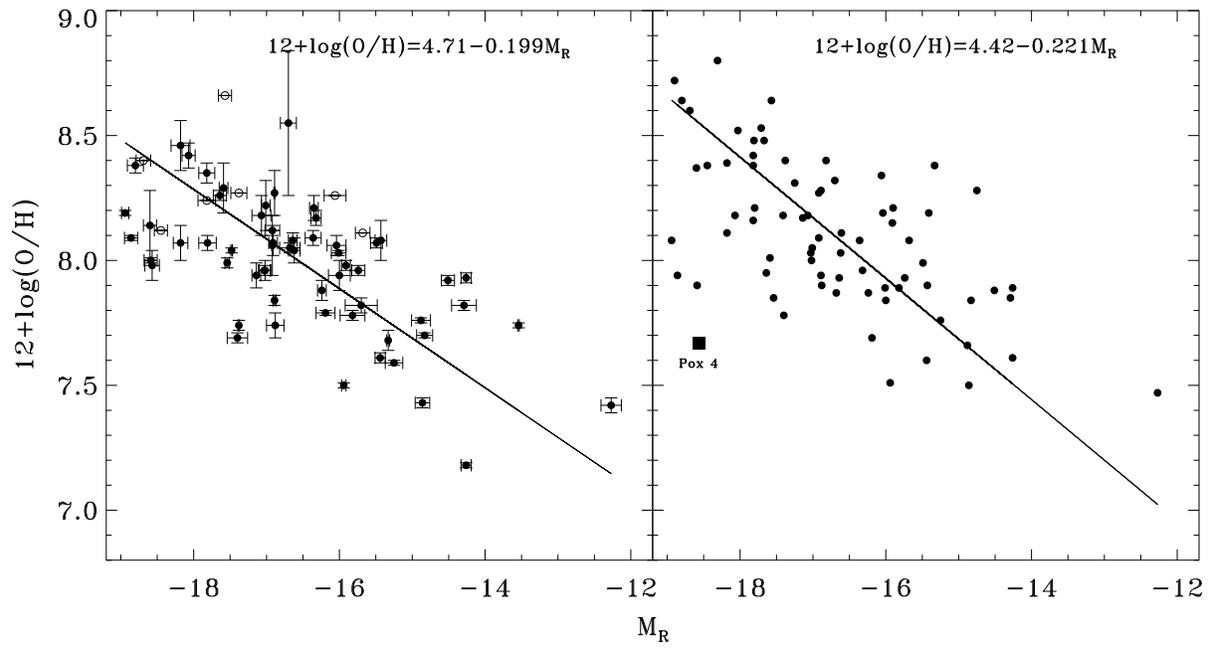}
\caption{Same as Fig. 2, but for the {\it R}-band.}
\label{Fig3}
\end{figure}

\newpage \clearpage
\begin{figure}
\centering
\includegraphics[width=\textwidth]{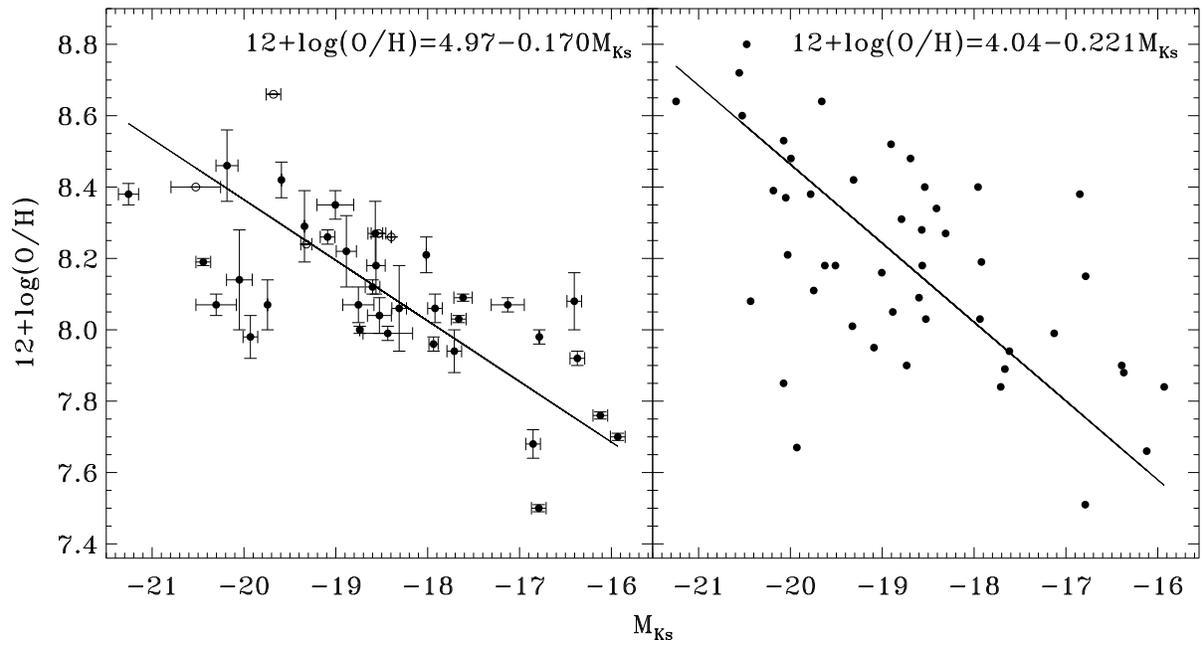}
\caption{Same as Fig. 2, but for the $K_s$-band}
\label{Fig4}
\end{figure}

\newpage
\clearpage
\begin{figure}
\centering
\includegraphics[width=\textwidth]{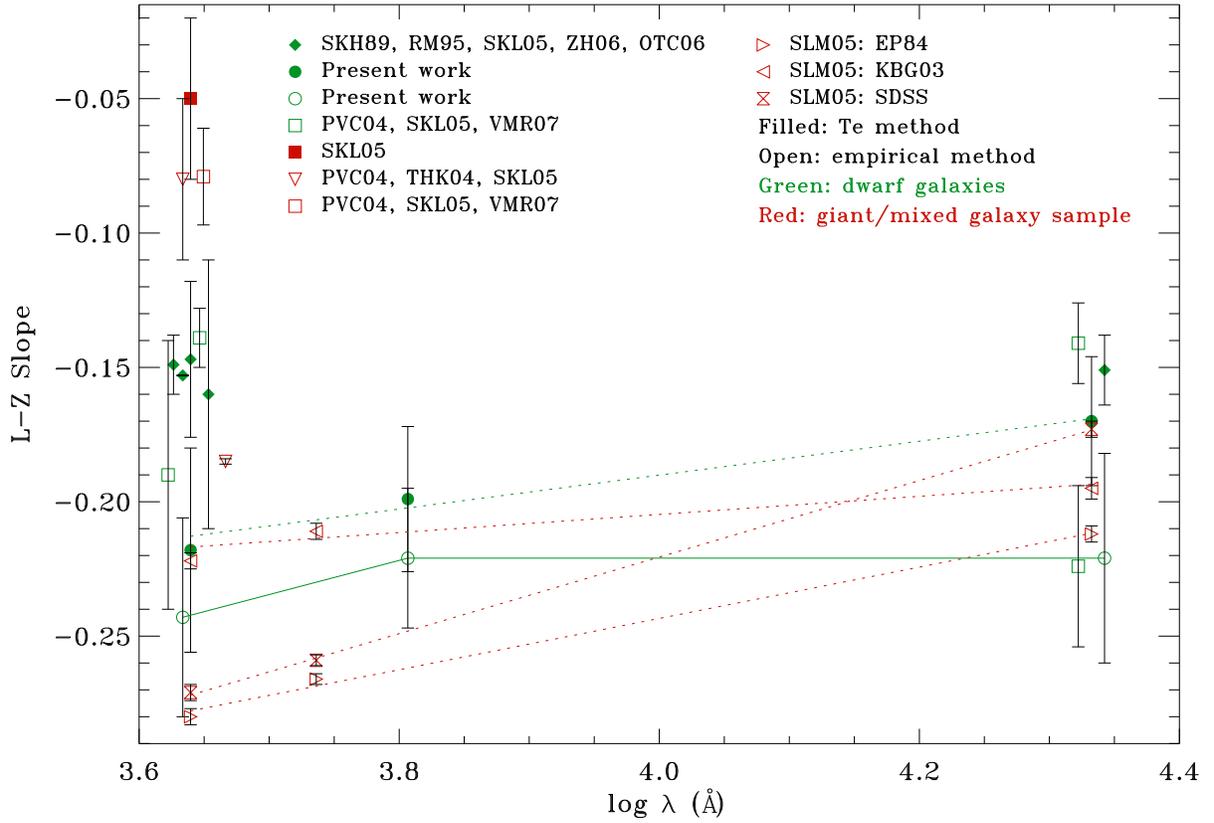}
\caption{Compiled slopes of \LZ\ relations. For clarity, some points for $B$- and $K_s$-band are shifted a bit along the x-axis. The dotted lines are the results of error-weighted linear fits to the data. The names of the references are the same as those in Table 4.}
\label{Fig5}
\end{figure}

\newpage \clearpage

\begin{figure}
\centering
\includegraphics[width=\textwidth]{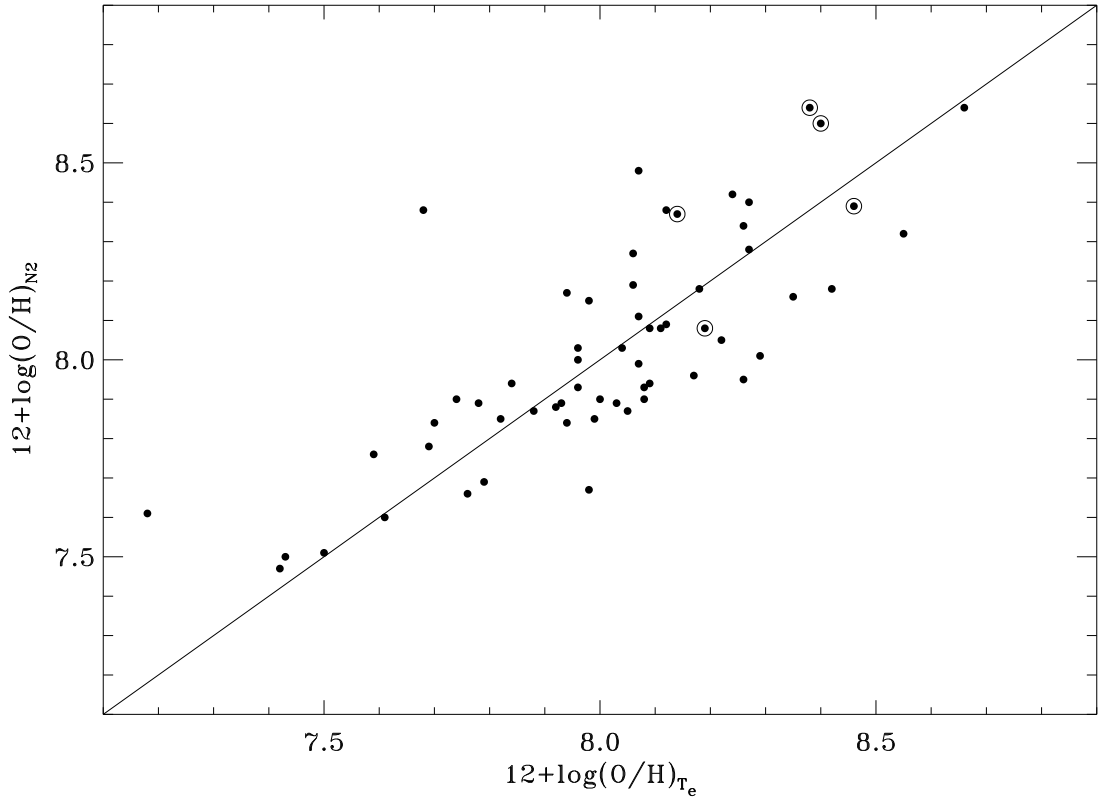}
\caption{Comparison of $T_e$-based (abscissa) and N2-based (ordinate) O/H. Points with circles represent galaxies with $M_{K_s} \leq -20.0$ mag. The solid line is a reference line for the case when the two quantities are the same.}
\label{Fig5}
\end{figure}

\newpage \clearpage
\begin{figure}
\centering
\includegraphics[width=\textwidth]{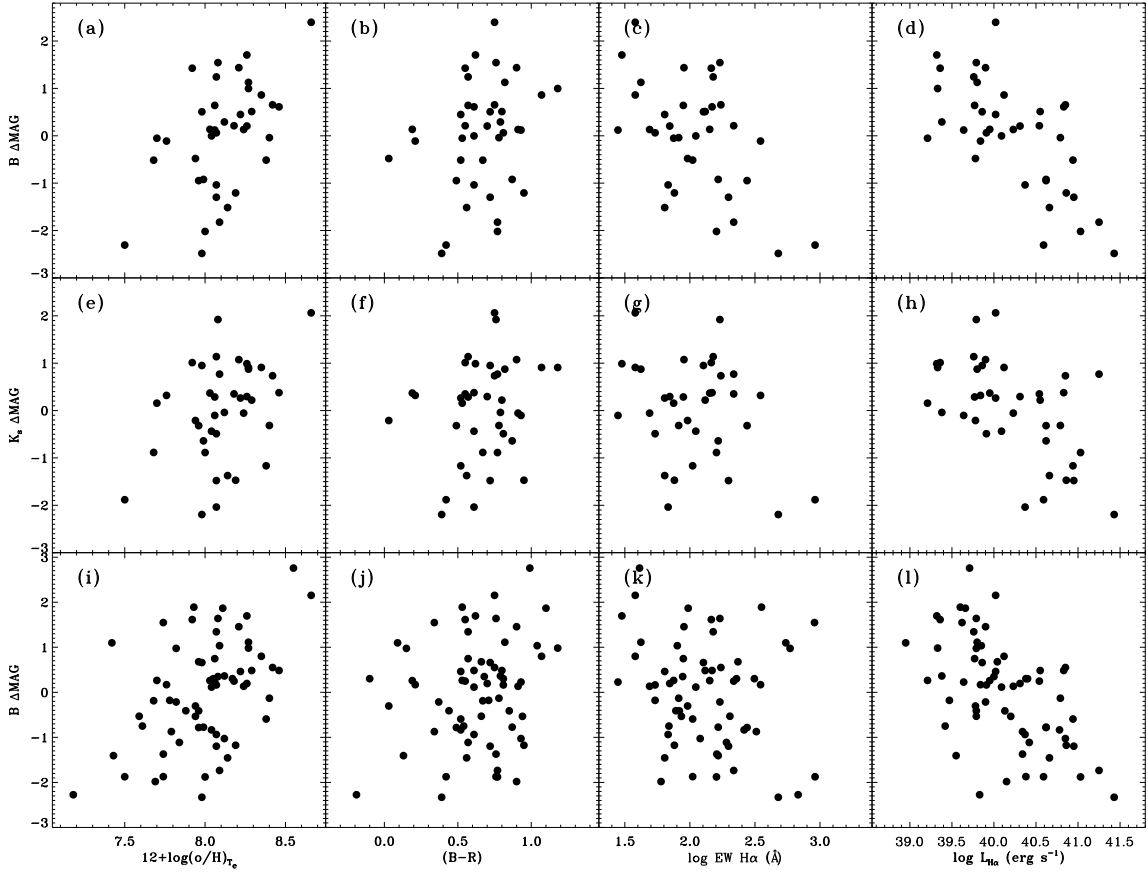}
\caption{Galaxy oxygen abundance, color, EW$_{{\rm{H}}\alpha}$ and H$\alpha$ luminosity (from the left to the right) vs. luminosity residuals, $\Delta M$, from the best-fit linear $T_e$-based \LZ\ relation. The bottom panels plot the full sample of a total of 66 BCDs, while the upper two panels plot the subsample of 39 galaxies which have $K_s$ magnitudes. Panels ({\it a}), ({\it e}) and ({\it i}) indicate that there is a weak correlation between $\Delta M$ and metallicity. While no correlation between $\Delta M$ and ($B-R$) color or EW$_{{\rm{H}}\alpha}$ can be investigated [see panels ({\it b}), ({\it f}), ({\it j}), ({\it g}) and ({\it k})]. By comparing panel ({\it c}) with ({\it k}), one can find that the correlation present in panel ({\it c}) is due to selection bias. This bias is also present in panel ({\it h}). Meanwhile, both of panels ({\it d}) and ({\it l}) show correlations between $\Delta M_B$ and $L_{{\rm{H}}\alpha}$, indicating that this correlation really exists.}
\label{Fig6}
\end{figure}

\end{document}